\documentclass[10pt,journal,compsoc]{IEEEtran}
\usepackage{setspace}
\ifCLASSINFOpdf
\usepackage[pdftex]{graphicx}
\else
\fi
\usepackage[cmex10]{amsmath}
\usepackage{array}
\begin{document}

%
\title{Performance analysis and Optimisation of the Met Unified Model on a Cray XC30}
%
%
%
%

\author{Karthee~Sivalingam, Grenville~Lister, Bryan~Lawrence 
\IEEEcompsocitemizethanks{\IEEEcompsocthanksitem K. Sivalingam,
B. Lawrence and G.Lister are with the The National Centre for Atmospheric Science
- Computational Modelling Services (NCAS-CMS), University of Reading, Reading, UK \protect\\
E-mail: k.sivalingam@reading.ac.uk
}
\thanks{Manuscript received ...; revised ..}}

\IEEEtitleabstractindextext{%
\begin{abstract}

The Unified Model (UM) code supports simulation of weather, climate and earth system 
processes. It is primarily developed by the UK Met Office, but in recent years a 
wider community of users and developers have grown around the code. Here we present 
results from the optimisation work carried out by the UK National Centre for Atmospheric 
Science (NCAS) for a high resolution configuration (N512 $\approx$ 25km) on the UK ARCHER 
supercomputer, a Cray XC-30.
On ARCHER, we use Cray Performance Analysis Tools (CrayPAT) to analyse the performance of UM and then
Cray Reveal to identify and parallelise serial loops using OpenMP directives.
We compare performance of the optimised version at a range of scales, and with a range of 
optimisations, including altered MPI rank placement, and addition of OpenMP directives. 
It is seen that improvements in MPI configuration yield performance improvements of 
between 5 and 12\%, and the added OpenMP directives yield an 
additional 5-16\% speedup. We also identify further code optimisations which could yield 
yet greater improvement in performance.  
We note that speedup gained using addition of OpenMP directives does not result in 
improved performance on the IBM Power platform where much of the code has been 
developed. This suggests that performance gains on future heterogeneous architectures 
will be hard to port. Nonetheless, it is clear that the investment of months in 
analysis and optimisation has yielded performance gains that correspond to the saving of tens of 
millions of core-hours on current climate projects.
\end{abstract}

}

\maketitle

\IEEEdisplaynontitleabstractindextext

%
\IEEEpeerreviewmaketitle

\section{Introduction}

The Unified Model (UM) is a simulation code which has been designed to support both 
predicting weather and projecting and understanding climate. It represents more than 
two decades of development and evolution by the UK Met Office and collaborators.
During this time regular upgrades added both improved science and better performance.  
The UM can be configured in a range of modes, from single-column through to global mode, 
and with a range of horizontal and vertical resolutions. The active use of this wide 
range of configurations is termed seamless prediction; the history of the evolution 
of the UM in this context is described in \cite{brown_unified_2012}.  While the UM is 
mainly used by the UK Met Office, it is increasingly used in other organisations, both 
in the UK, and elsewhere.  The UK academic community are one set of such users of the UM, 
with use of the UM underpinning a significant proportion of weather, and particularly, 
climate science.  Over the years the academic community have fed improvements in science 
back into the Met Office trunk, and occasionally they have provided performance 
improvements - generally those associated with migrating the code to new architectures. 
One such set of improvements is described here. We show how a combination of improving 
a range of MPI settings and the use of OpenMP directives can make speed ups of in 
excess of 20\% on a Cray XC-30 for some configurations of the UM. 

\begin{table}[!t]
\caption{Hardware specification of HPC machines}
\label{tab:hpc_spec}
\renewcommand{\arraystretch}{1.5}
\centering
\begin{tabular}{ l c c c}\hline
&\bfseries HECTOR & \bfseries ARCHER &\bfseries MONSooN \\\hline
Machine&Cray XE6& Cray XC30&IBM 775  \\
Compute node&Opteron&Ivybridge&Power 7 \\
Interconnect&Gemini&Aries&IBM\\
File system&Lustre&Lustre&GPFS\\
Compute Cores&90,112 & 118,080 & 5,120 \\
Memory (in TB) & 90  & 318.5  & 10 \\ \hline
\end{tabular}
\end{table}

\begin{table*}[!t]
\caption{Standard UM jobs at different resolution. The number of columns and rows describes 
the grid of the global model in North-South and East-West (horizontal) direction respectively. 
Land points refers to the number of simulated land points. Vertical levels describes the vertical 
grid of the atmosphere. Timesteps refers to the number of physics timesteps per simulated day.  
Resolution refers to resolution of the global grid. }
\label{tab:um_jobs}
\renewcommand{\arraystretch}{1.5}
\centering
\begin{tabular}{ c  p{1.5 cm} p{1.5 cm} p{2 cm} p{1.5 cm} p{1.5 cm} p{1.5 cm}}\hline
\bfseries Jobname &\bfseries Columns & \bfseries Rows &\bfseries Land Points  
&\bfseries Vertical levels &\bfseries Timesteps  &\bfseries Resolution \\\hline
N96    & 192   &144 &11271    & 85 & 20 min & 130 km \\
N216  & 432   &324 &52614    & 85 & 15 min & 60 km \\
N512  & 1024 &768 &280592 & 85 &10 min & 25 km \\ \hline
\end{tabular}
\end{table*}

The version of the UM discussed here is an evolution (GA6, V8.6) of the HadGEM3 global 
atmosphere configuration which includes a sophisticated land surface sub-model, 
JULES~\cite{walters_met_2014}. Most of the integration time is taken up by the atmospheric 
dynamical core. The original UM hydrostatic dynamical core of the UM described in
\cite{cullen_conservative_1991} was replaced in 2002 with the New Dynamics
~\cite{davies_new_2005}. The New Dynamics, a non-hydrostatic, semi-implicit, 
semi-Lagrangian scheme was used until late 2014 in operational numerical weather 
prediction and climate science configurations. Over the last year, variants of the 
model have progressively seen their dynamical cores replaced with the Even New 
Dynamics(ENDgame) scheme~\cite{ga6_inprep,end-game}. The ENDgame scheme itself is described 
in \cite{wood_inherently_2014}, with further scientific analysis of the performance and 
comparisons with the New Dynamics in \cite{mayne_using_2014}. The results presented here 
are the first analysis of the computational performance of an ENDgame climate configuration 
to appear in the open literature. 

In the UM, processes are discretized on a horizontal latitude-longitudinal grid, and over 
the years the grid-spacing has decreased, resulting in more simulated grid points 
and larger computational domains.  This march to higher resolution has enabled more 
complex process interactions, and significantly improved scientific outcomes. The 
code itself is mainly FORTRAN with a few calls to ANSI C routines, and can be run 
on any platform, but at anything except low resolution, requires a supercomputer.  
The primary versions of the shared UM code are currently maintained on an IBM Power 755
~\cite{quintero2014ibm} supercomputer at the Met Office (MONSooN, to be replaced 
by a Cray XC-40 in late 2015), and codes are typically ported from that environment 
onto target platforms.  In November 2011, the UM was ported to the HERMIT (Cray XE6) 
supercomputer at HLRS in Germany and optimised for the N512 high-resolution 
configuration ($\approx$25 km grid spacing) as part of a large simulation campaign 
(UPSCALE, \cite{gmd-7-1629-2014}).  Here we discuss a new port from the MONSooN environment
onto the ARCHER platform, which is aimed at extending and improving the UPSCALE optimisations.

ARCHER is a Cray XC-30, deployed at the Edinburgh Parallel Computing Centre (EPCC) as 
the national computing facility for environmental and engineering science. 
ARCHER has Intel Ivy-Bridge compute nodes~\cite{intel-micro}
and an Aries interconnect~\cite{cray-aries}. In what follows we examine performance 
of the UM on ARCHER at three different resolutions 
(N96, N216 and N512, corresponding to $\approx$ 130, 60 and 25 km) and compare with 
the MONSooN performance. We begin by discussing the default performance on MONSooN and 
ARCHER for these three configurations, then examine the impact of MPI communications on 
performance, leading to recommendations as to the best way of organising the layout of 
MPI ranks.  We then show that there is considerable scope for performance improvement 
if thread imbalance can be addressed, before exploiting the Cray Reveal tool to add OpenMP 
directives and get a significant speedup on ARCHER. We conclude by putting this work in 
context with similar work on previous versions of the UM as well as other similar codes, 
and making suggestions as to future optimisation potential.
%
%

Table \ref{tab:hpc_spec} lists the hardware specifications of the HPC machines that are
discussed in this paper. 
In this study, we use the IBM Power 775 machine as a baseline for comparison.
For MONSooN ,we assume that UM jobs are fully optimized by the MetOffice.
For ARCHER, optimizations discussed in the UPSCALE project 
are applied by default unless stated otherwise.
These jobs are used as a baseline for further performance analysis. 
The three resolutions of the UM models used are listed in table~\ref{tab:um_jobs} with the context 
discussed in ~\cite{mosac-um}.
We assume that the UM standard jobs will be used for performance analysis unless otherwise stated.

\section{Performance measurement}
In measuring the performance of UM models, we will use the number of model 
years simulated per day ($M_{year}$) as a metric.
\begin{equation}
M_{year} = \frac{1200}{T_{model} }
\end{equation}
where $T_{model}$ is the time taken for modelling 5 model days.
In this paper, a model year is assumed to be 360 days long and
$T_{model}$ is measured from the total wallclock time ($T_{wallclock}$ in seconds) and
the initial setup time ( $T_{initial}$ in seconds) of a 5 model day run.
\begin{equation}
T_{model} = T_{wallclock} - T_{initial}
\end{equation}

For global climate modelling, $M_{year}$ is the most useful metric and will be compared against
the number of physical cores $n_{core}$ used. The cost in core-hours of simulating a model
year ($C$) can be evaluated as follows
\begin{equation}
C = \frac{1}{M_{year}} \times n_{core} \times 24
\end{equation}
In this paper, $C$ will be scaled by 1/1000 for ease of representation and 1 $C$ will represent 1000 core-hours
or a kilo core-hours.

\subsection{Performance tuning parameters}

Bit reproducibility is a requirement for climate modelling and  is strictly enforced in all the runs. 
The following Cray FORTRAN compiler flags are used by default
\begin{verbatim}
    -e m -s real64 -s integer64 -h O2
    -hflex_mp=intolerant  -h omp
\end{verbatim}
to enforce bit reproducibility.
For all the UM jobs studied the default Lustre stripe count (4) and stripe size (1MB) are used.

For MPI parallelism, we can set the number of processes in the East-West (longitude) and 
North-South directions (latitude). 
The UM uses iterative solvers
to solve a Helmhotz equation and each iteration requires halos to be communicated between the MPI processes.
The interpolation order used in semi-lagrangian advection and the maximum wind speed allowed in the E-W direction
determines the size of the halo. The UM uses an extended halo size of
up to 8 and this restricts the maximum number of MPI processes in any direction. 
When the MPI processes are increased, the data columns/rows per MPI process reduces. 
This leads to overlapping halos that makes the UM model not to bit compare.

We can do an exhaustive search to find the optimal processor decomposition that can be used. 
This is very expensive and on the Cray XC30 we find that the peak performance of 
high resolution jobs has weak dependence on the decomposition. In further studies, we try to use 
a decomposition that is proportional to
the number of columns and rows of a job (as listed in table \ref{tab:um_jobs}).


OpenMP and IO server~\cite{Dennis01022012} support (asynchronous file IO) have been added in 
the recent versions. An IO server (Listener) puts the UM data writes in a FIFO queue and an IO server(Writer) 
processes the queue in an asynchronous manner. Parallelism is achieved by having multiple IO servers and using 
a threaded implementation for Listener and Writer.

For the UPSCALE project, IO servers are configured in dedicated node islands that are under populated
~\cite{gmd-7-1629-2014}. This requires many nodes to be dedicated to IO.
For MONSooN, the IO servers perform efficiently when they are spaced across 
the nodes running the UM~\cite{selwood-io}.
IO performance benchmarks on ARCHER show that the UM runs the fastest when all the
IO servers are placed on a single dedicated node.

There are many other parameters that can be tuned and are dependent on the resolution and how
the physics of the model is setup. Finding all the optimal setting for the UM jobs is beyond the scope of
this paper.
\begin{table}[!t]
\caption{Performance of Symmetric Multi-Threading (SMT) and Hyper-Threading technology (HT)
on MONSooN and ARCHER respectively. \%Speedup refers to the relative performance speedup
achieved by using SMT or HT.}
\label{tab:smt_ht}
\renewcommand{\arraystretch}{1.5}
\centering
\begin{tabular}{ c c c c }
\hline
\multicolumn{4}{c}{ARCHER} \\
\hline
& \multicolumn{2}{c}{Model years per day } & \\
\bfseries Cores &\bfseries HT OFF & \bfseries HT ON &\bfseries \%Speedup  \\ \hline
24 &    0.42&   0.48&   15.96\\
48 &	0.81&	0.90&	12.20\\
96 &	1.54&	1.74&	13.04\\
192&	2.80&	3.06&	9.44 \\
384&	4.78&	4.82&	0.80 \\
\hline
\multicolumn{4}{c}{MONSooN} \\
\hline
& \multicolumn{2}{c}{Model years per day } & \\
\bfseries Cores &\bfseries SMT OFF & \bfseries SMT ON &\bfseries \%Speedup  \\ \hline
32 &	0.69&	0.89&	29.30\\
64 &	1.31&	1.70&	30.26\\
128&	2.38&	3.09&	29.90\\
256&	4.10&	5.13&	25.21\\

\end{tabular}
\end{table}

\subsection{Threading}

In the UM, parallelisation is achieved through message passing and threads.
Symmetric multi threading (SMT) is supported in MONSooN and achieves significant speedup for the UM.
ARCHER (Intel Ivy bridge) supports hyper-threading(HT) which can be enabled by using the
'-j' option in aprun. '-j 2'  enables 2 hardware threads per PE as shown below.

\begin{verbatim}
    aprun -n 24 -j 2 UM.exe
\end{verbatim}

Enabling HT (or SMT) increases the number of processing elements (PEs) available per node. 
This enables the UM to run with twice the number of
MPI tasks or threads. Table \ref{tab:smt_ht} shows
the performance of HT and SMT on ARCHER and MONSooN respectively for a N96 job. The performance
is measured using only 2 threads per MPI task and the number of MPI tasks is doubled when SMT 
or HT is enabled.

While SMT shows a speedup of up to 30\% and shows good scaling up to 256 cores, 
the speedup achieved
from using HT is $\approx 16$\% but the speedup vanishes as the number of cores are increased to
384. For high resolution jobs that run on thousands of cores, we see that HT slows the UM.
So in all our model runs, HT is disabled and the UM is run with 12 MPI tasks and two threads
per node by default on ARCHER. For MONSooN, SMT is enabled and  32 MPI tasks and two threads
are used per node by default.

\begin{table}[!t]
\caption{Performance scaling of UM jobs on ARCHER. $EW$ refers to
number of PEs in East-West direction, $NS$ - number of PEs in North-South
direction, $T_{model}$ - Wallclock time taken to complete 5 model days,
$n_{core}$ - number of physical cores, $M_{year}$ - Model years simulated in a day,
$C$ - cost in core-hours per model year. }
\label{tab:archer_scaling}
\renewcommand{\arraystretch}{1.5}
\centering
\begin{tabular}{ c c c c c c c }
\hline
$EW$ & $NS$ &  Node & $T_{model}$  & $n_{core}$ & $M_{year}$ & $C$\\ \hline
\multicolumn{7}{c}{N96} \\
\hline
4&	3&	1&	2858&	24& 	0.42 & 1.38\\
4&	6&	2&	1476&	48&	    0.81 & 1.43\\
8&	6&	4&	771&	96&	    1.54 & 1.50\\
8&	12&	8&	416&	192&	2.80 & 1.65 \\
16&	12&	16&	237&	384&	4.78 & 1.93\\
24&	16&	32&	144&	768&	7.14 & 2.59\\
\hline
\multicolumn{7}{c}{N216} \\
\hline
12&	8&	8&	 2395&	192&	0.50 & 9.20\\
12&	16&	17&	1239&	408&	0.97 & 10.11\\
24&	16&	33&	685&	792&	1.75 & 10.85\\
24&	32&	65&	407&	1560&	2.95 & 12.70\\
36&	32&	97&	305&	2328&	3.93 & 14.20\\
48&	32&	129& 252&	3096&	4.76 & 15.60\\
48&	40&	161& 224&	3864&	5.36 & 17.31\\
\hline
\multicolumn{7}{c}{N512} \\
\hline
36&	24&	73&	2218&	1752&	0.54 & 77.72\\
36&	36&	109&	1617&	2616&	0.74 & 84.60\\
48&	36&	145& 1263&	3480&	0.95 & 87.90\\
48&	48&	193&	1025&	4632&	1.17 & 94.96\\
60&	48&	241&	885&	5784&	1.36 & 102.38\\
72&	50&	301&	773&	7224&	1.55 & 111.68\\
\hline
\end{tabular}
\end{table}

\begin{table}[!t]
\caption{Performance scaling of UM jobs on MONSooN. $EW$ refers to
number of PEs in East-West direction, $NS$ - number of PEs in North-South
direction, $T_{model}$ - Wallclock time taken to complete 5 model days,
$n_{core}$ - number of physical cores, $M_{year}$ - Model years simulated in a day,
$C$ - cost in core-hours per model year.}
\label{tab:monsoon_scaling}
\renewcommand{\arraystretch}{1.5}
\centering
\begin{tabular}{ c c c c c c c }
\hline
$EW$ & $NS$ &  Node & $T_{model}$  & $n_{core}$ & $M_{year}$ & $C$\\ \hline
\multicolumn{7}{c}{N96} \\
\hline
4&	8&	1&	1300&	32  &	0.89 &0.86\\
8&	8&	2&	696 &	64  &	1.70 &0.90\\
8&	16&	4&	377 &	128 &	3.09 &0.99\\
16&	16&	8&	218 &	256 &	5.13 &1.20\\
\hline
\multicolumn{7}{c}{N216} \\
\hline
4&	16&	2&	4010&	64  &	0.30& 5.13\\
8&	16&	4&	2117&	128 &	0.57& 5.42\\
8&	32&	8&	1140&	256 &	1.05& 5.84\\
16&	32&	16&	625 &	512 &	1.92& 6.40\\
32&	32&	32&	392 &	1024&	3.06& 8.03\\
\hline
\multicolumn{7}{c}{N512} \\
\hline
16&	32&	17&	4270&	544 &	0.28& 46.46\\
32&	32&	33&	2425&	1056&	0.49& 51.22\\
32&	48&	49&	1793&	1568&	0.67& 56.23\\
50&	40&	64&	1500&	2032&   0.8 & 60.96\\
\hline
\end{tabular}
\end{table}

\section{UM Performance}

Tables \ref{tab:archer_scaling} and \ref{tab:monsoon_scaling} show the performance scaling of the UM
jobs at three different resolutions on ARCHER and MONSooN respectively.
Performance is measured as number of model years simulated in a day ($M_{year}$) and the cost in
core-hours per model year( $C$). 

\begin{figure}[!t]
\centering
\includegraphics[width=3.6in]{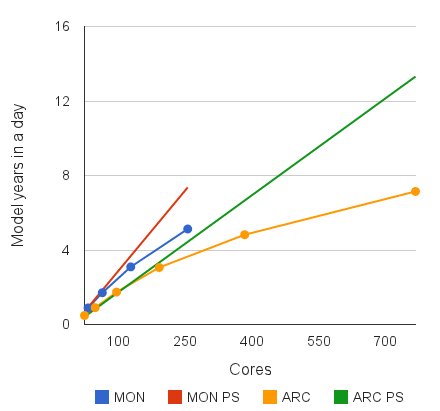}
\caption{Performance scaling of the N96 job on ARCHER(ARC) and MONSooN(MON). Cores refers to
the actual number of physical cores used and performance is measured as number of model years
simulated in a day ($M_{year}$). MON PS and ARC PS refers to perfect scaling that can be expected on MONSooN 
and ARCHER respectively.}
\label{fig_n96}
\end{figure}

\subsection{N96}

Figure~\ref{fig_n96} shows the scaling of the N96 job on ARCHER (ARC) and MONSooN (MON).
ARCHER perfect scaling (ARC PS) and MONSooN perfect scaling (MON PS) is plotted for reference
and refer to the respective perfect scaling that can be expected based on UM
performance on a single node ( or lowest number of nodes).

The UM scales to $M_{year} \approx 3$ with 128 and 192 cores on MONSooN and ARCHER
respectively. We can infer from this that the IBM Power 7 cores are 1.5
times faster than the Intel Ivy bridge cores.

On MONSooN, the N96 job scales only up to 256 cores as extended halo size restricts the number
of MPI PEs in any direction. 
Further scaling can be obtained by running 
the UM underpopulated (i.e.less than 32 MPI tasks per node). On ARCHER, the UM scales
to 768 cores. Using 3 times the number of cores as on MONSooN,
ARCHER has a peak performance that is 1.4 times than that on MONSooN.

\begin{figure}[!t]
\centering
\includegraphics[width=3.6in]{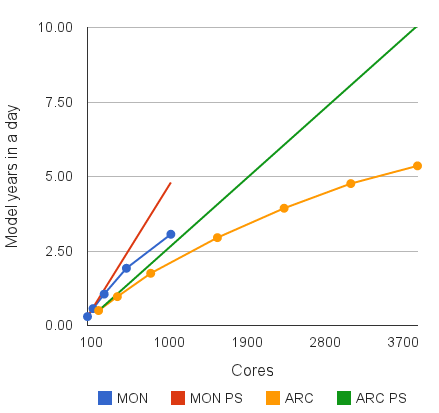}
\caption{Performance scaling of the N216 job on ARCHER(ARC) and MONSooN(MON). Cores refers to
the actual number of physical cores used and performance is measured as number of model years
simulated in a day ($M_{year}$). MON PS and ARC PS refers to perfect scaling that can be expected on MONSooN 
and ARCHER respectively.}
\label{fig_n216}
\end{figure}

\subsection{N216}
Figure~\ref{fig_n216} shows the scaling of the N216 job on ARCHER (ARC) and MONSooN (MON).
N216 shows similar scaling performance to that of N96. Using $\approx 3.8$ times the number
of cores as on MONSooN, ARCHER has a peak performance that is 1.75 times than that on MONSooN.
On ARCHER, IO servers are used to hide the overheads due to IO. 
On MONSooN N216 scales only up to 32 nodes (1024 cores) and IO servers do not improve the
performance at this scale.

Though the N216 scales to 3824 cores on ARCHER, the performance reduces almost by half 
compared to the perfect scaling.
This implies that the cost of simulating a model year ($C$) doubles as the  number of cores is
increased from 192 to 3864.

\begin{figure}[!t]
\centering
\includegraphics[width=3.69in]{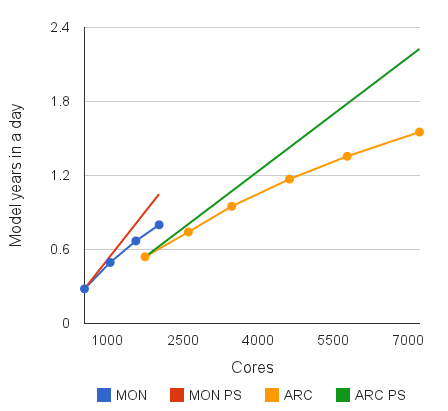}
\caption{Performance scaling of the N512 job on ARCHER(ARC) and MONSooN(MON). Cores refers to
the actual number of physical cores used and performance is measured as number of model years
simulated in a day ($M_{year}$). MON PS and ARC PS refers to perfect scaling that can be expected on MONSooN 
and ARCHER respectively.}
\label{fig_n512}
\end{figure}

\subsection{N512}
Figure~\ref{fig_n512} shows the scaling of the N512 job on ARCHER (ARC) and MONSooN (MON).
IO servers are used both on ARCHER and MONSooN. Using $\approx 3.55$ times the number of cores
as on MONSooN, ARCHER has a peak performance that is $\approx 1.94$ times than that on MONSooN.
On ARCHER ,the cost of simulating a model year ($C$) increases 30\% as the  number of cores is
increased from 1752 to 7224. On MONSooN the cost of simulating a model year
($C$) increases 23\% as the number of cores is increased from 544 to 2552.

\begin{figure*}[!t]
\centering
\includegraphics[width=7in]{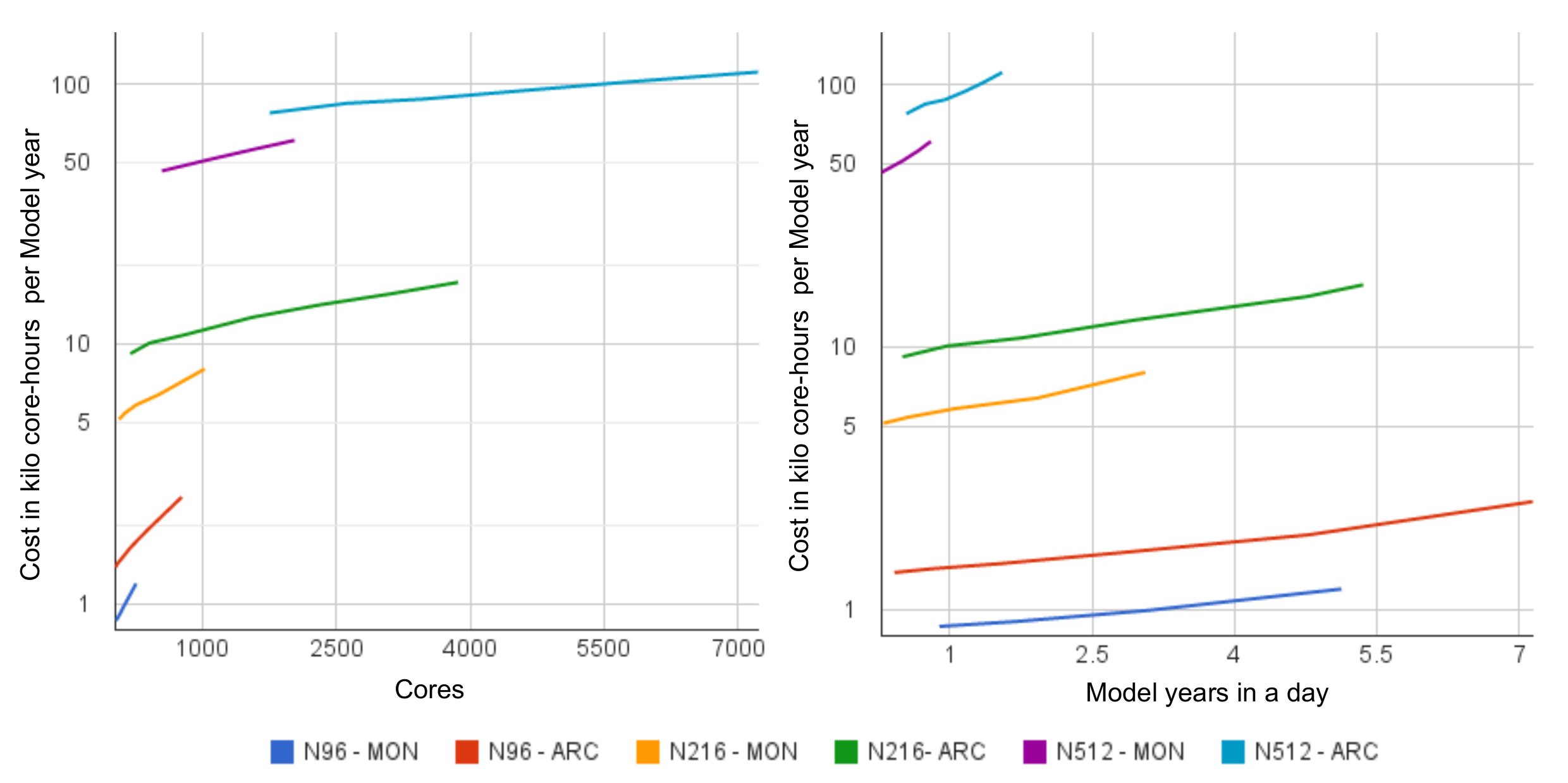}
\caption{Cost of simulating a model year ($C$) of the UM job on ARCHER(ARC) and MONSooN(MON) compared
to the number of physical cores - $n_{core}$ (left) and Model years in a day  $M_{year}$ (right).   }
\label{fig_cost}
\end{figure*}

Figure~\ref{fig_cost} compares the cost of simulating a model year ($C$) of the UM jobs on 
ARCHER(ARC) and MONSooN(MON) to the number of physical cores ($n_{core}$) and model years simulated 
in a day ($M_{year}$). In this figure, a horizontal line represents perfect scaling as this represents a constant cost
 of simulating a model year ($C$) as $n_{core}$ or $M_{year}$ is increased. Also a longer line represents
the better performance scaling.  $n_{core}$ scales well as the resolution is increased. We are more interested in $M_{year}$ and we can infer directly that the cost of jobs scales poorly with increase in resolution.
We see a general trend where MONSooN is cost efficient when compared to ARCHER, but
ARCHER scales better than MONSooN. This is based on the assumption that the usage of IBM power 7 core
cost the same as the Ivy bridge core.
On ARCHER N512 achieves only 1.55 model years in a day whereas we can
model 5 and 7 model years of N216 and N96 respectively. Further comparing the peak performance of these jobs 
xon ARCHER, N512 is 6.5 times costlier than N216 and 43 times costlier than N96 jobs. This clearly shows the need 
to analyze and optimize the high resolution N512 model.

\begin{figure*}[!t]
\centering
\includegraphics[width=6.5in]{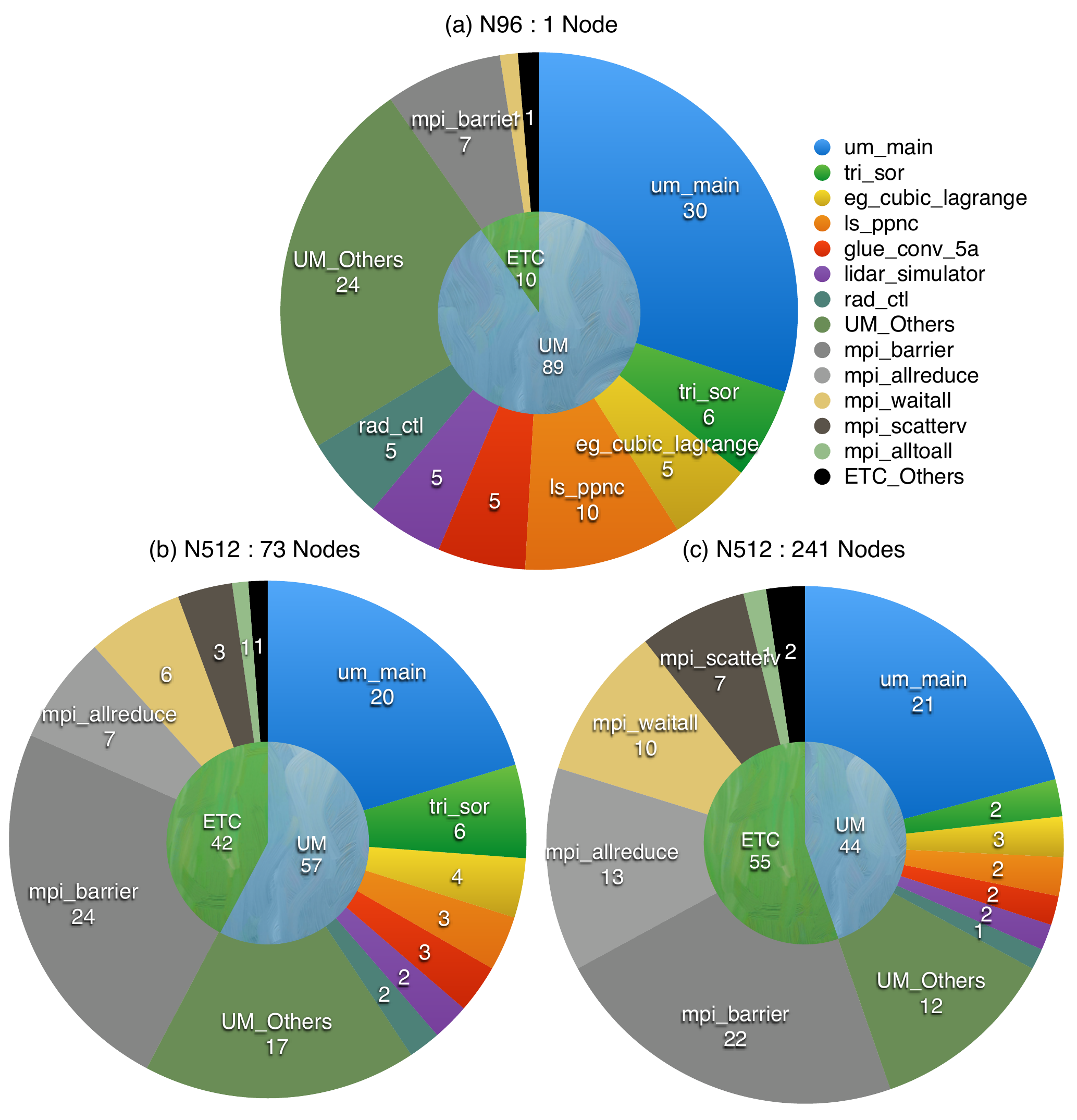}
\caption{Pie chart showing profile of UM jobs. (a) N96 job running on a single node.
(b) N512 job running on 73 nodes (c) N512 job running on 241 nodes. The inner pie chart shows the
overall profile in which UM includes profile of all UM user code and ETC includes all other library calls.}
\label{fig_profile}
\end{figure*}

\section{Performance Analysis}
Profiling the application is the first step in analyzing the performance and understanding the
bottlenecks. Cray Performance Analysis Tool (CrayPAT)~\cite{cray-pat} is used to profile the
UM on ARCHER. Using CrayPAT, we can also profile a specific user defined function such as IO, SHMEM and more.
For the UM which has a flat profile, we use the automated program analysis. This analyses the UM
performance and identifies interesting areas/functions that should be instrumented.

In this analysis, we will use the N96 job running on a single node ($N96_1$) as a baseline
and compare it with the profile
of N512 job, running on 73 ($N512_{73}$) and 241 ($N512_{241}$) nodes.
This will help us understand the bottlenecks
of a high resolution job compared to a lower resolution job and
also understand how the profile changes as N512 is scaled from 73
to 241 nodes.

Figure~\ref{fig_profile} shows the profile of the $N96_1$, $N512_{73}$ and $N512_{241}$ jobs in
the form of pie charts. The inner pie charts show the summary of the profile and the outer pie charts
reveal the finer details of the exact functions/procedures.
In the profiles, UM refers to the profile of UM user code
(like um\_main, tri\_sor) and ETC to that of all library calls (like mpi\_barrier, mpi\_alltoall).
UM\_Others/ETC\_Others includes all other UM/ETC instrumented functions that do not
individually consume significant execution time.

The profiles are based on a 5 day model run. The initial setup and other related
overheads are included in 'um\_main'. Hence this does not include the actual simulation time.
In long climate runs, this overhead ('um\_main') becomes negligible.

We can infer from the fugure~\ref{fig_profile} that the UM has a flat profile as even the most expensive functions 
(excluding um\_main) consume less than 10\% of
the profile. The profiles of UM jobs change with resolution as the most expensive function for
$N96_1$, $N512_{73}$ and $N512_{241}$ is not the same.

\begin{figure*}[!t]
\centering
\includegraphics[width=5.5in]{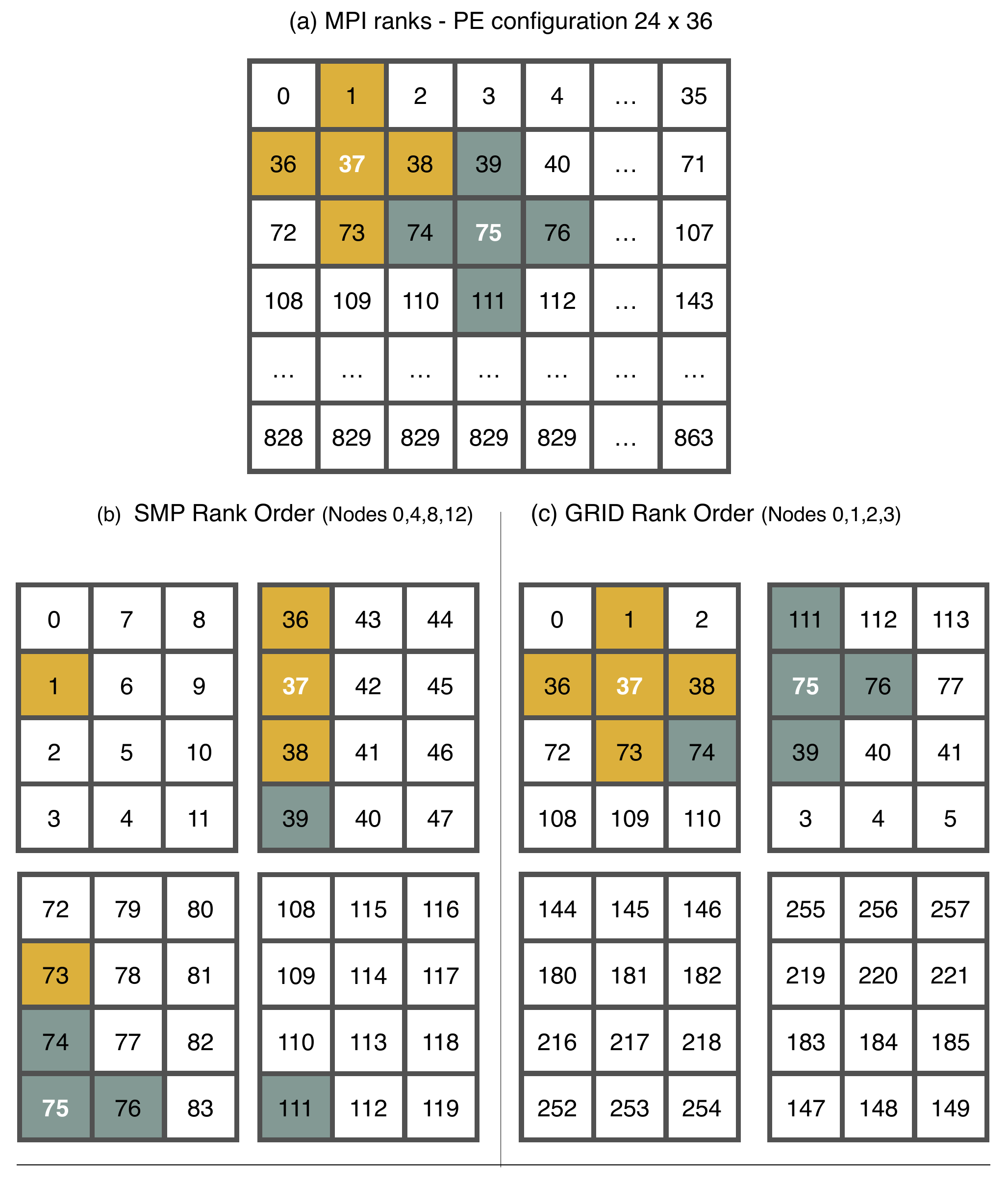}
\caption{(a) MPI ranks of a UM job with 24$\times$36 PE configuration.
(b) Placement of MPI ranks using the default SMP rank order.
(c) Placement of MPI ranks using the GRID rank order.
MPI rank 37 (and 75) along with the ranks involved in nearest neighbor communications
are highlighted. Rank order is based on using 12 MPI ranks per node on ARCHER.}
\label{fig_pe_decomp}
\end{figure*}

\subsection{MPI rank reorder}

The $N96_1$ job runs on a single node and does not include any off node communication.
The N512 jobs run on more than 72 nodes and require message passing between nodes.
For ARCHER, the off node communications are significantly more expensive
when compared to intra node communications.

The ETC profile represents the message passing overheads and increases from 10\% on $N96_1$ to 42\%
on $N512_{73}$. For N512, ETC increases by 13\% as the number of nodes is
increased from 73 to 241. Also ETC consumes more the half of the total CPU time when scaled
to 241 nodes. We can deduce that message passing between nodes is the most significant bottleneck 
for scaling of high resolution models. 

CrayPAT can be configured to detect the MPI grid and the MPI communication pattern used. Based on this
it suggests several MPI rank orders that will reduce off node traffic and increase
the MPI bandwidth. On ARCHER the MPI ranks can be reordered by
setting the environment variable MPICH\_RANK\_REORDER\_METHOD to 3 and specifying the
rank order in MPICH\_RANK\_ORDER file.

MPICH\_RANK\_REORDER\_METHOD is set by default to 2, which is symmetric multiprocessing
(SMP) style placement. For the N512 job, CrayPAT suggests a different grid order that is based on
nearest neighbor communications.
This rank order is generated using a utility called grid\_order. For example, a N512 job running
with $24(EW) \times 36(NS)$ PE decomposition, CrayPAT recommends the rank order generated by the
following command.

\begin{multline}
grid\_order \, -R \, -P\, -c \,4,1 \,-g \,24,36 \\
-m\, 864 \,-n \,12\, -N\, 12
\end{multline}

Here -R refers to row-major order, -m the maximum rank count, and -N the number of
ranks per node. Refer to \cite{cray-pat}  for the details of the options used.
We will refer to this rank reorder as GRID.

Figure~\ref{fig_pe_decomp} shows MPI ranks for UM jobs with $24(EW)\times36(NS)$ PE configuration
and how the MPI ranks can be placed on ARCHER nodes in SMP and GRID style. We assume that 12 MPI ranks
will be placed per node as used in all our UM jobs. To illustrate the message passing,
we can consider the nearest neighbor
communications of ranks 37 and 75 (as highlighted).

For SMP, 4 nodes are involved in the
communication whereas the number of nodes is reduced to 2 for GRID.  We are more 
interested in the 
off-node communications as they are costlier than inter-node communications. In SMP, rank
37 has to communicate with ranks 1 and 73 which are off-node. In GRID, all the nearest neighbors
of 37 reside on the same node. Similarly for rank 75, the number of MPI ranks that are off-node
are reduced by half if GRID rank order is used instead of SMP .

\begin{figure}[!t]
\centering
\includegraphics[width=3.59in]{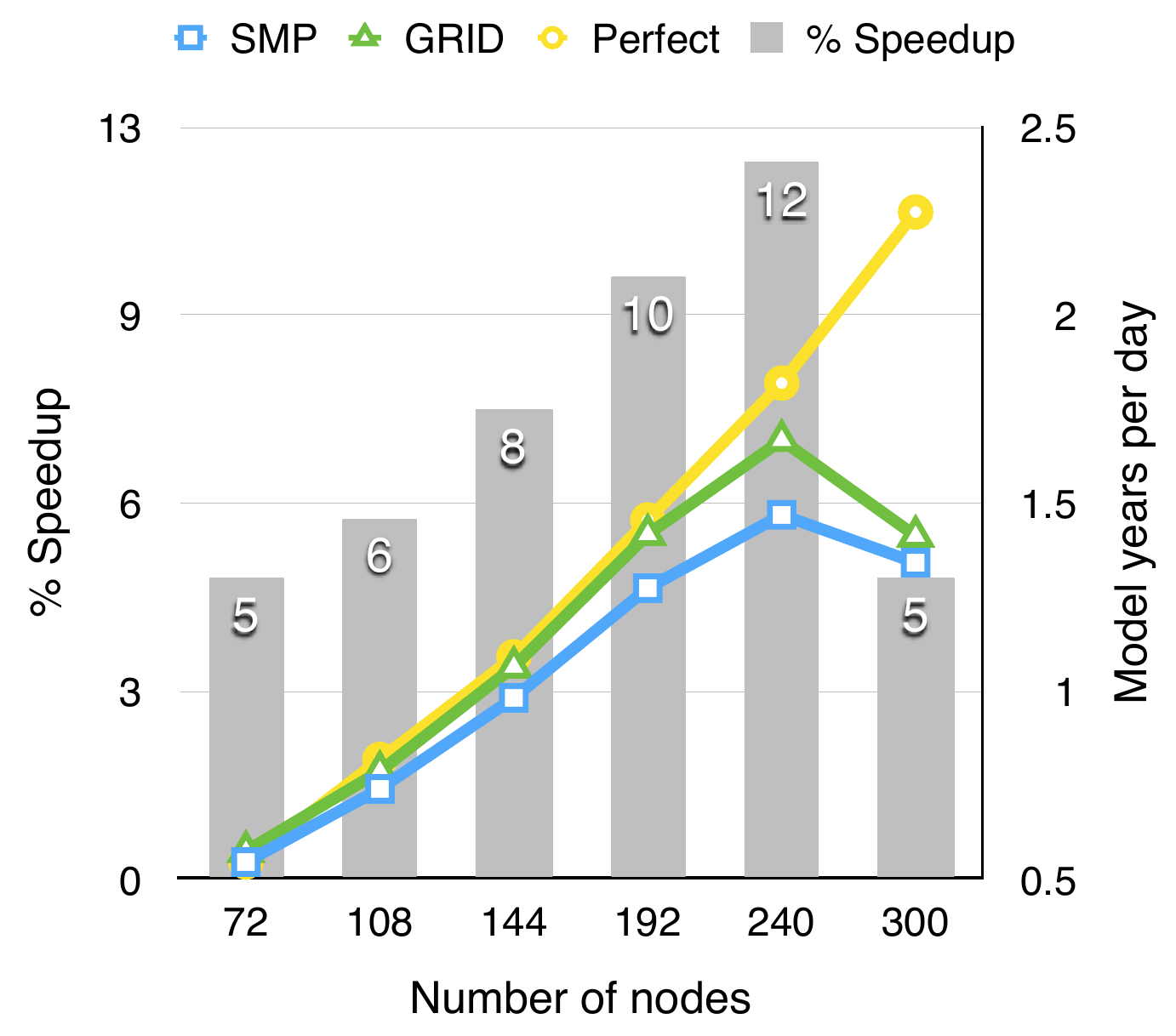}
\caption{Comparison of performance scaling of the N512 job with different MPI rank orders. \%Speedup
refers to the relative speedup achieved by using GRID rank order instead of SMP rank order.}
\label{fig_mpi_tuning}
\end{figure}

Figure~\ref{fig_mpi_tuning} compares the performance of a N512 job using a SMP and GRID rank
order. GRID achieve a speedup up to 12\% compared to the SMP rank order. GRID rank order
results in almost a perfect speedup when scaled up to 192 nodes. In performing these measurements,
IO is turned OFF. The IO performance is dependent on the Lustre file system which is a
shared resource. In measuring speedup of the order of 10\%, the measurements become unreliable as
the shared file system performance is noisy .

\subsection{Load imbalance}

Another important metric is the load balance of the application. CrayPAT reports on the
load balance information over all PEs and threads. It also provides finer details
of load balance of different functions/routines. 
In CrayPAT, imbalance is measured as
imbalance percentages which are relative to the set of threads or PEs.
For example, if we consider the UM running with 2 OpenMP threads, an imbalance percentage of 50\% implies
that one thread is idle for 50\% of the time when the other thread is busy.

\begin{figure*}[!t]
\centering
\includegraphics[width=6.7in]{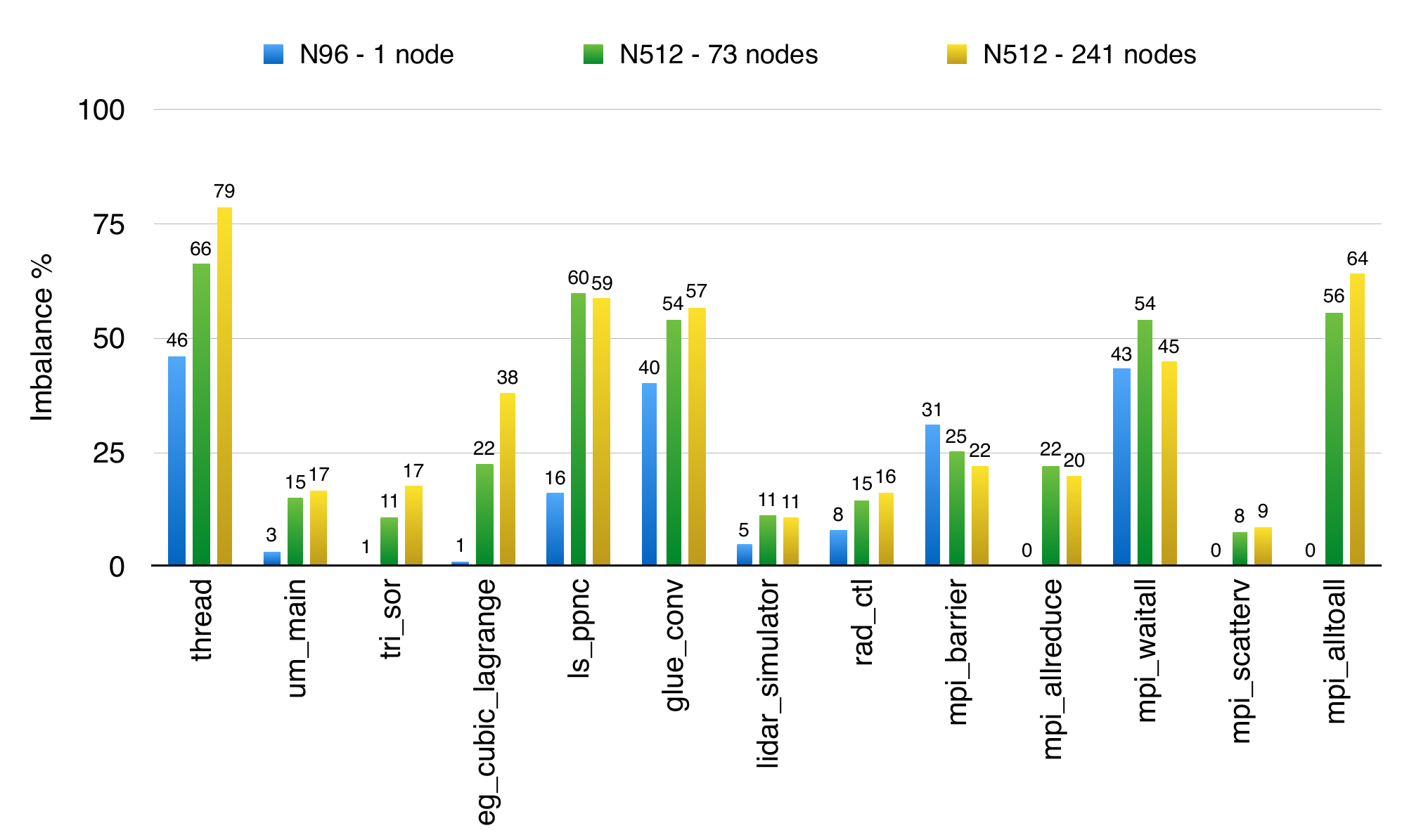}
\caption{Imbalance percentage of UM jobs : N96 job running on a single node;
N512 job running on 73 nodes; N512 job running on 241 nodes; Thread imbalance percentage are relative
to the set of threads used. Imbalance percentage of UM functions are relative to set of PEs. }
\label{fig_load_balance}
\end{figure*}

Figure~\ref{fig_load_balance} shows the imbalance percentage of
$N96_1$, $N512_{73}$ and $N512_{241}$. All these measurements are based on a 5 day model run
using 12 MPI tasks and 2 OpenMP threads per node. 
For $N96_1$, 'mpi\_allreduce', 'mpi\_scatterv' and 'mpi\_alltoall' have no imbalance but increases up to
64\% for $N512_{241}$. This further emphasizes the need to reduce message passing overheads.

In SMP based supercomputers,  we need thread based parallelism to improve the scaling of the UM and 
reduce MPI overheads which increase to more than 50\% of the profile as the N512 model is scaled to 241 nodes.
For example, using n threads will allow the UM to scale to n times the number of cores as that of a MPI only version
and also reduce the number of MPI packets communicated between nodes by half. 
The number of MPI tasks in the UM is limited by the extended halo size.
Also the extended halos increase the actual memory consumed. . 
This clearly shows the need for better thread performance to improve the scaling of high resolution models.

On ARCHER, if we assume that the OpenMP implementation is 100\% efficient and scales well, 
we can ideally set the number of threads to 12 and MPI tasks to 2.  ARCHER nodes have two, 12-core NUMA 
regions and the 12 cores of a single NUMA region have fast access to the shared local memory. 
$N96_1$ has only 10\% MPI overhead and the 46\% thread imbalance of  $N96_1$ can be attributed directly to
loops that are not thread parallelised. So based on Amdahl's law, we can expect poor scaling of the UM as the 
number of threads is increased. Functions 'glue\_conv' and 'ls\_ppnc' have higher imbalance compared to 
other UM functions and can be improved by parallelising loops.

\section{OpenMP Optimisation }
OpenMP provides a standard and portable way of parallelising loops. This involves scoping the
loop variable (as shared, private ...) and inserting OpenMP directives before a loop. 
The UM has a flat profile with thousand of serial loops that can be parallelised. 
Parallelising all the loops is expensive and
careful consideration is required to ensure data consistency. Also race conditions involving
parallel threads are hard to debug.

Cray Reveal is an integrated performance analysis and code optimisation tool. It
provides loop analysis and scoping of serial loops and
suggests OpenMP directives that can be inserted to a loop.
Performance data collected during execution by CrayPAT can easily be attached to Cray Reveal
to identify the profile of loops.
This will help us in prioritizing the loops that consume more CPU time.

The tool also shows the compiler optimizations that have been applied.
Even though this tool is user friendly, it requires knowledge of OpenMP to
resolve conflicts, race conditions and scoping issues.
It works only with the Cray compiling environment. This tool
does not provide support for parallel regions, task based parallelism, barrier, 
critical or atomic regions.

\begin{table}[!t]
\caption{Performance scaling of UM jobs using increasing number of OpenMP threads on ARCHER. 
Wallclock time refers to the time taken to complete 2 model days. 
MPI$\times$THR refers to the number of MPI tasks  $\times$ number of OpenMP threads used per node.
$UM_{8.6}$ refers to the original UM code and $UM_{Reveal}$ to the code with new OpenMP directives.
\%Speedup is measured as a relative performance improvement achieved by adding new OpenMP directives. }
\label{tab:cray_reveal_a}
\renewcommand{\arraystretch}{1.5}
\centering
\begin{tabular}{ c c  c c  c }
\hline
& & \multicolumn{2}{c}{\bfseries $M_{year} \, with  \, $} &  \\
\bfseries Node & \bfseries MPI$\times$THR &\bfseries $UM_{8.6}$  
& \bfseries $UM_{Reveal}$  &\bfseries \%Speedup \\ \hline
\multicolumn{5}{c}{N96 (PE config - 4$\times3$)}   \\
\hline
1& 12$\times$2 &	0.408&	0.429&	4.9  \\
2& 6$\times$4  &	0.568 &	0.625 &	9.1  \\
3& 4$\times$6  &	0.577 & 	0.712 &	18.9 \\
\hline
\multicolumn{5}{c}{N512 (PE config - 36$\times24$)}  \\
\hline
72&12$\times$2&	0.494&	0.518&	4.6\\
144&6$\times$4&	0.667&	0.731&	8.9\\
216&4$\times$6&	0.734&	0.872&	15.9\\
\hline
\end{tabular}
\end{table}

\subsection{UM - Reveal on ARCHER}
UM has a flat profile and has thousands of loops that can be parallelised. Parallelising all these
loops is beyond the scope of this paper. As a case study, 2389 serial loops are parallelised by
adding OpenMP directives as suggested by Cray Reveal.
default(none) option is used for all the newly added directives. For
Fortran array notation expressions like 'for all' and 'where' statements, are parallelised using
the following directives
\begin{verbatim}
    !$OMP PARALLEL WORKSHARE
    !$OMP END PARALLEL WORKSHARE
\end{verbatim}
Bit reproducibility is enforced to test for correctness of OpenMP directives.

Table \ref{tab:cray_reveal_a} shows the performance scaling of UM jobs when the number of threads
is increased on ARCHER. In all the performance measurements, each thread is assigned to a PE which means
PEs are not oversubscribed. For example, in case of ARCHER nodes, when the number of OpenMP 
threads is increased from 2 to 4, the number of MPI tasks per node is reduced from 12 to 6.

In this section, the original UM code will be
referred to as $UM_{8.6}$ (8.6 is the original UM version number used)
and the UM code with Cray Reveal changes will be referred to as
$UM_{Reveal}$. Hyperthreading and symmetric multi-threading are switched off for ARCHER and MONSooN
respectively, to enable easier comparison.

On ARCHER, $UM_{Reveal}$ achieves a speedup of 18.9\% and 15.9\% for N96 
and N512 jobs respectively with 6 OpenMP threads. The fall  in speedup for N512
job compared to N96, is due to additional MPI overheads.  For both N96 and N512 jobs, the speedup increases by
more than 3 times as the number of threads are trebled. This is a significant improvement 
and can be further improved by adding OpenMP directives to thousands of other serial loops.

\begin{table}
\caption{Performance scaling of a N96 job (PE config - 4$\times$4) using increasing number of 
OpenMP threads on MONSooN. 
Refer to table~\ref{tab:cray_reveal_a} for definition of 
Wallclock time, MPI$\times$THR and  \%Speedup.
}
\label{tab:cray_reveal_m}
\renewcommand{\arraystretch}{1.5}
\centering
\begin{tabular}{ c c  c c  c }
\hline
& & \multicolumn{2}{c}{\bfseries $M_{year} \, using $} &  \\
\bfseries Node & \bfseries MPI$\times$THR &\bfseries $UM_{8.6}$  & \bfseries $UM_{Reveal}$  
&\bfseries \%Speedup \\ \hline
1& 16$\times$2 &  0.276 & 0.277 & 0.7 \\
2& 8$\times$4 &	0.370 & 0.373 & 0.9\\
4& 4$\times$8 &	0.427 & 0.374 & -14.3\\
\hline
\end{tabular}
\end{table}

\begin{figure*}[!t]
\centering
\includegraphics[width=6.5in]{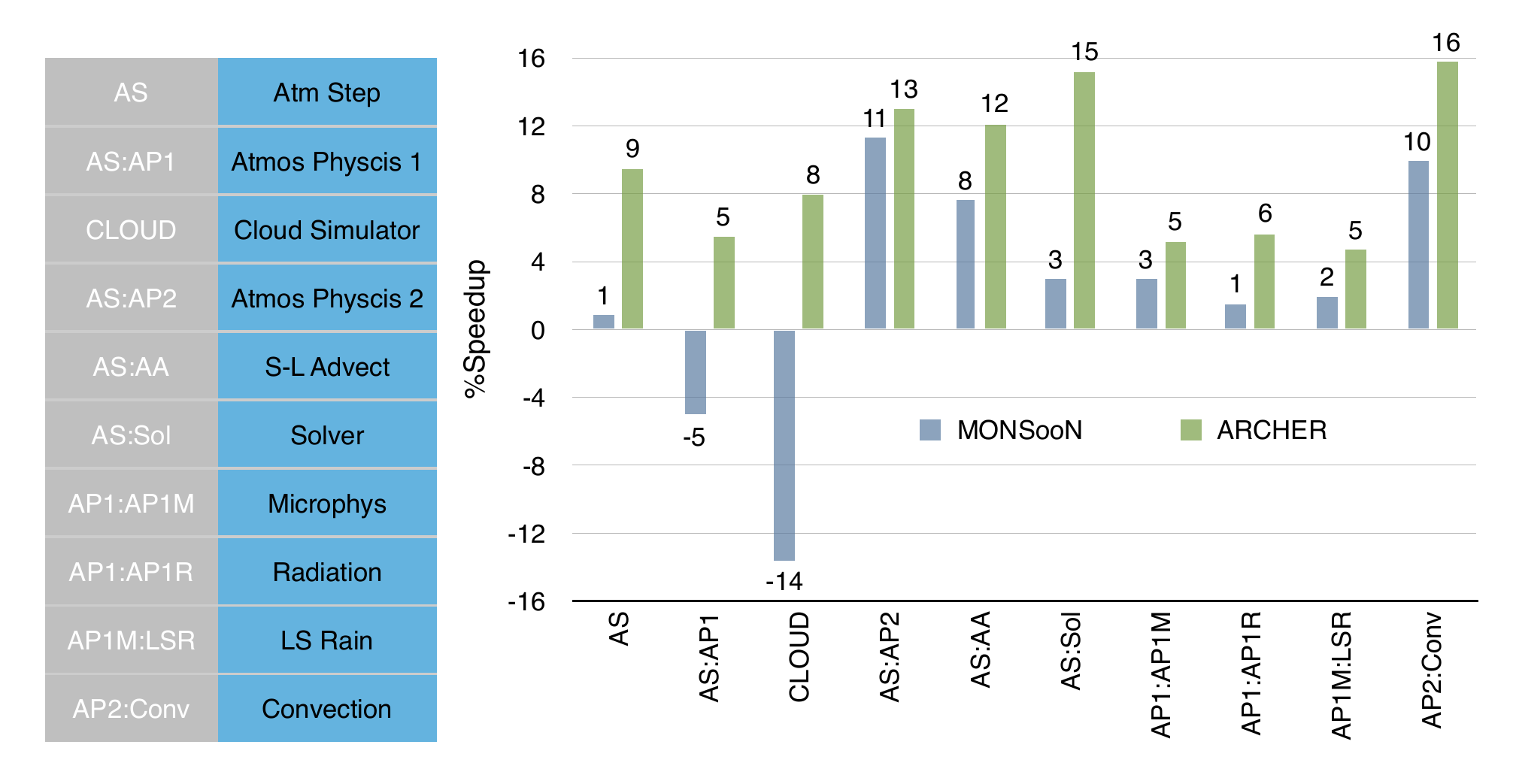}
\caption{\%Speedup of UM functions on MONSooN and ARCHER. \%Speedup
is measured as the relative improvement achieved by adding OpenMP regions. Performance of
UM is measured using 6 MPI and 4 OpenMP threads per node. In X-axis labels, the calltree of the function is 
specified using ':' delimiter. For example in AS:AP1, 'Atm Step'  calls function
'Atmos Physics 1'.}
\label{fig_cray_reveal}
\end{figure*}

\subsection{UM - Reveal on MONSooN}

$UM_{Reveal}$ is based on the OpenMP standard and can be ported easily to most
other supercomputers. The IBM compiler supports OpenMP and $UM_{Reveal}$ is easily ported to MONSooN.
Table \ref{tab:cray_reveal_m} shows the performance scaling of UM jobs when the number of threads
is increased on MONSooN.
On MONSooN, $UM_{Reveal}$ does not result in significant speedup
and when 8 threads are used, $UM_{Reveal}$ slows the performance by 14.3\%.

Figure~\ref{fig_cray_reveal} shows a closer look at the relative speedup achieved
for different functions on MONSooN and ARCHER using 4 OpenMP threads. \%Speedup
is measured as the relative improvement achieved by using $UM_{Reveal}$ instead of $UM_{8.6}$.
In ARCHER, all the UM functions show considerable speedup whereas on MONSooN,
'Atmos Physics 1' and 'Cloud Simulator' routines slow down considerably.

Even though OpenMP provides a standard and portable way for implementing thread based parallelism,
the performance improvements are not portable and depend on the hardware architecture.
One major difference we found is
that ARCHER has a unified L3 cache whereas MONSooN does not.
This may be the reason for the significant
difference in performance. Since MONSooN is being replaced by a Cray XC40 machine,
this is not investigated further.

\subsection{ Related work }

The performance of the UM is dependent not only on threads and MPI processes, but also on the resolution, the physics
and the algorithms used. This requires careful modelling of UM performance to study the impact of all these
contributing factors. Researchers have modelled the performance
of the UM to study scalability with increased resolution and core counts~\cite{Kerbyson:2005,annette-poster}.
Historically the UM has been ported by the Met Office to different architectures, for example Cray T3E, NEC SX-6 , 
NEC SX-8, IBM Power 6 and Power 7~\cite{selwood-arch}. They studied components that 
affect the scalability and found IO and the Helmhotz solver to be the biggest obstacles to UM scaling.
Also these components interact with each other to a different extent and this makes model 
performance evaluation difficult.

Researchers at the Australian National University 
have developed efficient profiling methodology and scalability analysis of the UM at different
resolutions~\cite{6008986}. They have identified that the high resolution N512 L70 job 
scales up to 2048 cores but is affected by load imbalance and MPI communications. 
On an Intel Sandy Bridge based cluster, they identified 
that hybrid OpenMP/MPI will provide the best opportunity for optimisations. These findings are based on 
UM version 7.5 (New Dynamics) released in 2010 and we see similar trends as discussed in this paper. 

Met Office researchers have studied the scaling of high resolution jobs on a IBM Power 6 and an  Intel
Nehalem cluster (Juropa). They found that the scaling can be improved by using IO servers, OpenMP 
and identified further scope for improvement~\cite{um-scalability} .
This study reports  $\approx 6 \%$ speedup achieved using 2 OpenMP threads and SMT.  Also the GungHo
project~\cite{gungho} which is research collaboration between the Met Office, NERC funded researchers 
and STFC Daresbury strongly supports introducing thread based parallelism to improve scalability by 
reducing the cost of MPI communications. Also they emphasise the need to exploit thread parallelism 
that will be available in future exascale machines. The OpenMP optimisations we have discussed in this paper
make good strides in improving the thread parallelism of UM.

As part of the UPSCALE project, Tom Edwards at Cray studied the performance
of the UM (version 8.0, N512 L80, New Dynamics) on HERMIT, a Cray XE6 machine~\cite{tedwards-upscale}. 
In his report, the UM was also optimised by 
reordeing the ranks so that the IO servers are placed in separate dedicated nodes. This and other optimisations
resulted in 39\% improvement in runtime performance of the UM for 14\% increase in number of cores.  This
reports also suggests the need to extend OpenMP based thread parallelism which severely affects the scaling 
of the UM. 

In \cite{Mercier}, the authors  prescribe
that the MPI ranks should be reordered to match the application communication pattern with the underlying hardware. Also \cite{Brandfass2013372} found MPI reordering to be a promising optimisation for unstructured CFD code.
In \cite{reading34184}, the researchers studied the performance of a simple shallow water
model on a Cray XE6 machine (HECTOR). They suggest that MPI ranks should be mapped
to physical cores such that the off-node data transfer volume in a nearest neighbour communication
can be reduced. This is in agreement with the our findings based on the performance of the UM on ARCHER, 
a Cray XC30 machine. 

Similar efforts have been made to study the performance of other weather models. Weather and Research
Forecasting model (WRF) is used all around the world and its performance has been studied in detail on 
Blue Gene/L~\cite{Kerbyson07}, Blue Gene/P~\cite{Malakar:2012}, Cray XE6~\cite{Johnsen:2013},
Cray XT~\cite{ifs-crayxt} and many other machines. 
ECMWF ( European Centre for Medium-Range Weather Forecasts) 
have ported IFS (Integrated Forecasting System) from IBM Power 7 to Cray XC30~\cite{ifs-cray} 
and identified performance testing to be an important tool in meeting application challenges. As a part of 
ECMWF's scalability project, IFS has been ported to GPU, Intel Xeon Phi and Intel Xeon Haswell
and its performance compared~\cite{ifs-future}.

\section{Conclusion}
We have analysed of the performance of the UM at three different resolutions
on ARCHER using CrayPAT tools and compared to that of MONSooN.
Even though IBM Power 7 cores are more powerful compared to Intel Ivy bridge,
ARCHER scales better compared to MONSooN. Performance analysis 
shows that MPI communication and thread imbalance affect the scaling of high resolution
UM jobs. Reordering the MPI ranks using GRID rank order speed up the UM by up to 12\%
compared to default SMP rank order. Using Cray Reveal, new OpenMP directives
are added to the UM that results in improved speedup of up to 16\% on ARCHER 
whereas it slows the performance on MONSooN. These performance optimisations have resulted in 
savings of tens of millions of core-hours in current climate projects.
We can further improve the UM performance by allowing both threads to make MPI calls and by adding OpenMP 
directives to loops that can be parallelised. 

For high resolution (N512) jobs, thread imbalance increases from 66\% to
79\% as the number of nodes is increased from 73 to 241(as shown in  figure~\ref{fig_load_balance}). 
This is because only thread0 is involved in message passing. The imbalance can be reduced
by allowing both threads to make MPI calls. This requires not only using a thread-safe
MPI implementation~\cite{Gropp:2006}, but also careful
coding to ensure data consistency and avoid race conditions. 
Exploring this beyond the scope of this paper.
Even though we parallelised 2389 serial loops using Cray Reveal, there are thousands of loops
that can still be parallelised. Also parrallel regions can be added to reduce threading overheads.
This will further improve the scaling and efficiency of the UM performance.


%

\ifCLASSOPTIONcompsoc
\section*{Acknowledgments}
\else
\section*{Acknowledgment}
\fi

The work done here was funded by a contract to the National Center for Atmospheric Science - 
Computational and Modelling Services group (NCAS-CMS) at the University of Reading by the 
Met Office via the Joint Weather and Climate Research Programme (JWCRP). We are grateful to 
colleagues at the Met Office and at CMS for assistance and advice throughout this project, 
in particular Matthew Mizielinski, Rosalyn Hatcher and Simon Wilson. We thank EPCC for their
help in porting and supporting the UM jobs on ARCHER.

\ifCLASSOPTIONcaptionsoff
\newpage
\fi




\bibliographystyle{IEEEtran}
\bibliography{IEEEabrv,papers}

\vfill




\end{document}